\documentclass[12pt]{article}
\usepackage{amsmath,amssymb,amsthm,mathtools}
\usepackage{graphicx,psfrag,epsf,tikz}
\usepackage{enumerate}
\usepackage[round]{natbib}
\usepackage{setspace}
\usepackage{algorithm, algpseudocode}
\usepackage{color}
\usepackage[title]{appendix}

\newtheorem{theorem}{Theorem}[section]

\newtheorem{definition}{Definition}

\newcommand{\blind}{0}
\newcommand{\bsy}{\boldsymbol}
\newcommand{\mbf}{\mathbf}
\newcommand{\mbb}{\mathbb}
\newcommand{\mcl}{\mathcal}

\newcommand{\hrmt}{^H}
\newcommand{\noop}[1]{}

\begin{document}

\title{\LARGE\bf Marginally Parametrized Spatio-Temporal Models and Stepwise Maximum Likelihood Estimation}
\author{Matthew Edwards,\thanks{School of Mathematics, Statistics and Physics, Newcastle University, Newcastle upon Tyne, NE1 7RU, United Kingdom (E-mail: \textit{m.edwards3@ncl.ac.uk}).}\quad Stefano Castruccio
\thanks{Department of Applied and Computational Mathematics and Statistics, University of Notre Dame, Notre Dame, IN 46556, United States of America (E-mail: \textit{scastruc@nd.edu}).}\quad and Dorit Hammerling\thanks{Department of Analytics and Integrative Machine Learning, National Center for Atmospheric Research, Boulder, CO 80307, United States of America (E-mail: \textit{dorith@ucar.edu}).}}

\def\spacingset#1{\renewcommand{\baselinestretch}%
{#1}\small\normalsize} \spacingset{1}

\if0\blind
{
   \maketitle
} \fi

\if1\blind
{
  \bigskip
  \bigskip
  \bigskip
  \begin{center}
    {\LARGE\bf Marginally Parametrized Spatio-Temporal Models and Stepwise Maximum Likelihood Estimation}
	\end{center}
  \medskip
} \fi

\bigskip
\begin{abstract}
In order to learn the complex features of large spatio-temporal data, models with large parameter sets are often required. However, estimating a large number of parameters is often infeasible due to the computational and memory costs of maximum likelihood estimation (MLE). We introduce the class of marginally parametrized (MP) models, where inference can be performed efficiently with a sequence of marginal (estimated) likelihood functions via stepwise maximum likelihood estimation (SMLE). We provide the conditions under which the stepwise estimators are consistent, and we prove that this class of models includes the diagonal vector autoregressive moving average model. We demonstrate that the parameters of this model can be obtained at least three orders of magnitude faster using SMLE compared to MLE, with only a small loss in statistical efficiency. We apply a MP model to a spatio-temporal global climate data set (in order to learn complex features of interest to climate scientists) consisting of over five million data points, and we demonstrate how estimation can be performed in less than an hour on a laptop.
\end{abstract}

\vfill

\spacingset{1.45}

\newpage
\section{\label{sec:intro}Introduction}
The ever-increasing availability of large and complex spatio-temporal data in many areas of science including census, global climate, neuroscience and epidemiology, calls for the development of flexible models able to capture the features of interest to stakeholders. A model flexible enough to learn structural dependence at multiple scales, however, comes at the cost of large parameter sets required to describe it, and hence makes traditional inferential approaches such as maximum likelihood estimation (MLE) often infeasible. MLE depends on both the computational and memory costs of evaluating the likelihood function and the number of evaluations of the optimization algorithm, and while the costs of each evaluation increase with the number of data points, the number of evaluations increases with the number of parameters. It is therefore essential to consider the costs of evaluation and the number of evaluations when specifying models with large parameter sets for large complex spatio-temporal data.
   
    For a Gaussian model, if the data set consists of $N$ data points and the covariance matrix is unstructured, then these costs are $\mcl{O}(N^3)$ floating point operations (flops) and $\mcl{O}(N^2)$ memory (in the non-Gaussian case, the cost is typically considerably higher: a max-stable likelihood evaluation requires $\mcl{O}(B_N)$ flops, where $B_N$ is the Bell number of order $N$, see \cite{cas16ms}). Approaches that reduce these costs often use structured covariance matrices (e.g~sparse, low-rank, circulant) \citep{golub2012matrix,davis79circulant} to leverage numerical linear algebra routines (e.g.~FFT) and parallel computation. For large spatio-temporal data, approaches include: weighted composite likelihoods \citep{bevilacqua2012estimating}, full-scale approximations \citep{zhang2015full}, dynamic nearest neighbour processes \citep{datta2016nonseparable} and dynamic multi-resolution spatial models \citep{johannesson2007dynamic}. In certain cases, approaches for large spatial data can be applied to the innovations of dynamic models (e.g.~VARMA models). These approaches include: fixed rank kriging \citep{cressie2008fixed}, lattice kriging \citep{nychka2015multiresolution}, predictive processes \citep{banerjee2008gaussian}, covariance tapering \citep{kaufman2008covariance}, multi-resolution approximations \citep{katzfuss2017multi}, nearest neighbour processes \citep{datta2016hierarchical} and stochastic partial differential equations \citep{lindgren2011explicit}. 

	All these approaches focus on models (possibly misspecified, in the case of composite likelihoods) with likelihood  functions that have low costs of evaluation, hence their applicability to large spatio-temporal data. While these methods have been proven to be scalable for a wide range of applications, the ability to learn complex spatio-temporal patterns could come at the price of large parameter sets, and hence it is essential to consider the number of evaluations as well as the costs of evaluation. The number of evaluations depends on the number of iterations of the optimization algorithm (e.g.~two evaluations per iteration for a numerical derivative) which, in turn, depends on the number of parameters $P$ \citep{vavasis1991nonlinear}. If the likelihood function is concave then algorithms exist where the number of iterations is polynomial in $P$ \citep[e.g.~ellipsoid algorithm,][]{yudin1977evaluation}. However, if the likelihood function is non-concave, algorithms \textit{only} exist where the number of iterations is exponential in $P$ \citep{nemirovskii1983problem}. Therefore, MLE is infeasible for models with large parameter sets if the likelihood function is non-concave. In order to estimate a large number of parameters, parameter subsets must be estimated in multiple steps via a stepwise estimation method.
    
    Multi-stage approaches are stepwise estimation methods often used in applications with different types of dependence (e.g.~temporal, spatial) such as non-linear mixed models \citep{giltinan1995nonlinear}. \citet[pp.~431-433]{schabenberger2017statistical} proposed a two-stage approach where time series submodels are fit at each spatial location and then a spatial model is fit to the combined residuals. A large class of multi-stage approaches has been proposed in the context of global climate data \citep{castruccio2013global,cas17} where estimation in time, longitude and latitude is performed in three stages. An extension has been proposed in \cite{cas16} where estimation of altitude dependence is performed in a fourth stage. In the context of neuroscience a multi-stage approach for whole-brain data was proposed in \citep{cas18biom}. These classes of models have a reduced evaluation cost, as each submodel is defined over fewer data points. They also require a reduced number of evaluations, as each submodel depends on fewer parameters. Therefore, they are ideal for dealing with large and complex spatio-temporal data.
   
    These multi-stage methodologies have emerged in the aforementioned statistical literature as a collection of \textit{ad-hoc} methods for large complex spatio-temporal data where the nature of the problem suggests that modeling at multiple resolutions is useful for learning properties of interest. The lack of an underlying and unifying framework has so far prevented a full understanding of the generalizability of such models. Perhaps more importantly, a multi-stage approach does not guarantee an underlying joint model, and hence does not always allow probabilistic statements about the data. Without a joint model, minimum mean squared error prediction cannot be performed as conditional models cannot be derived. This is particularly problematic in spatio-temporal statistics where prediction is often the primary objective, but it is also a concern in simulation as without a joint model there is no explicit distribution to sample. Furthermore, the lack of a proper class of models did not allow for providing any asymptotic consistency results. 

	The aim of this work is to provide the foundation for multi-stage approaches by defining a joint model that can then be applied to large complex spatio-temporal data. This very general class of models is termed marginally parametrized (MP), does not require any Gaussian or stationary assumption, and is such that multi-stage inference is always possible with submodels that are proper marginal distributions. We define the general stepwise estimation method, termed stepwise maximum likelihood estimation (SMLE), and we provide the conditions that a MP model must satisfy to achieve asymptotic consistency across all stages of the inference.
    
	The remainder of this paper is structured as follows: in Section~\ref{sec:model} we define the class of MP models and a spatio-temporal model that is a member of this class, i.e.~the diagonal VARMA model. In Section~\ref{sec:inference}, we introduce the SMLE method, the asymptotic consistency theorem and the application of SMLE to the diagonal vector autoregressive moving average (VARMA) model. In Section~\ref{sec:sim} we provide two simulation studies. The first compares MLE with SMLE for the diagonal VARMA model with Mat\'{e}rn covariance \citep{stein2012interpolation} and the second corroborates the consistency theorem for the corresponding estimators. In Section~\ref{sec:app} we apply the diagonal VARMA model to a large complex spatio-temporal global climate data set. We conclude in Section~\ref{sec:con}.

\section{\label{sec:model}Marginally Parametrized Model}
	Here we introduce the class of marginally parametrized (MP) models and a spatio-temporal model that is a member of this class, i.e.~the diagonal VARMA model.

	\subsection{Definition and Heuristics}
Denote $\mbf{y}$ a data set consisting of $N$ data points and denote $\mcl{L}(\mbf{y}\mid\bsy\theta)$ a corresponding joint likelihood function that depends on a parameter vector $\bsy\theta$ consisting of $P$ parameters.

\begin{definition}[Marginally Parametrized Models]
\label{def:MP}
A model for $\mbf{y}$ is MP if there exists a finite sequence of $K>1$ data subsets $(\mbf{y}_k)$ such that the marginal model of $\mathbf{y}_k$ depends on a parameter subset with a partition $\bsy\theta_k,\bsy\eta_k$ where $\bsy\theta_k\neq\varnothing$ and $\bsy\eta_k\subseteq\bsy\theta_1\cup\dots\cup\bsy\theta_{k-1}$ ($\bsy\eta_1=\varnothing$) for $k=1,\dots,K$ and $\bsy\theta_1,\dots,\bsy\theta_K$ is a partition of $\bsy\theta$.
\end{definition}
	
    The sequence of data subsets $(\mathbf{y}_k)$ corresponds to a sequence of marginal models. Each marginal model depends on a parameter subset that is partitioned into a set of primary and nuisance parameters ($\bsy\theta_k$ and $\bsy\eta_k$ respectively). The primary parameters correspond to the parameters previous marginal models in the sequence \textit{do not} depend on; whereas, the nuisance parameters correspond to the parameters previous models in the sequence \textit{do} depend on. Heuristically, the primary parameters of each marginal model control the dependence \textit{only} within data subsets; hence, marginally parameterized. As a consequence of these conditions, the parameter set of a MP model can be estimated with a sequence of marginal (estimated) likelihood functions \citep[Section~10.6]{pawitan2001all}, see Subsection~\ref{ssec:SMLE}. 

	\subsection{\label{ssec:dVARMA}Diagonal VARMA Model}
We now consider a special case where $\mbf{y}$ is a spatio-temporal data set where the sampling design consists of $T$ regularly spaced time points $t\in\mbb{Z}_{>0}$ at $S$ arbitrarily spaced, but fixed in time, locations $\mbf{x}_s\in\mcl{M}^d$, where $\mcl{M}^d$ is a $d$-dimensional manifold (e.g.~plane, sphere); for a total of $N=S\cdot T$ data points. Denote $\mbf{Y}_t$ the random vector corresponding to the data points at all the locations at time point $t$. The diagonal VARMA model \citep{lutkepohl2005new} with autoregressive (AR) order $p$ and moving average (MA) order $q$ is defined as
\begin{equation}
\label{eq:dVARMA}
\mbf{Y}_t=\bsy\mu+\Sigma\mbf{U}_t+\sum_{i=1}^{p}\Phi_i\mbf{Y}_{t-i}+\sum_{j=1}^q\Pi_j\Sigma\mbf{U}_{t-j}\quad\text{where}\quad\mathbf{U}_t\overset{\text{i.i.d.}}{\sim}\mcl{N}(\mbf{0},\text{R}(\bsy\nu)),
\end{equation}
where $\bsy\mu=(\mu_1,\dots,\mu_S)$ is the vector of mean parameters, $\Sigma=\text{diag}(\sigma_s)$ is the diagonal matrix of standard deviation parameters, $\Phi_i=\text{diag}(\phi_{i,s})$ are the diagonal matrices of AR parameters and $\Pi_j=\text{diag}(\pi_{j,s})$ are the diagonal matrices of MA parameters. Furthermore, $\mbf{U}_t$ are i.i.d.~in time, centered and unscaled Gaussian innovations and $\bsy\nu$ is the set of correlation parameters. This is a very flexible model: the mean, standard deviation, AR and MA parameters can be different at each location and the correlation matrix $\text{R}$ has no constraints.

	To prove that \eqref{eq:dVARMA} is a MP model we must demonstrate that there exists a finite sequence of data subsets that satisfy the conditions of Definition~\ref{def:MP}. Consider the finite sequence of $K=S+1$ data subsets $(\mbf{y}_k)$ where $\mbf{y}_k$ is the time series at spatial location $\mbf{x}_k$ for $k$ from 1 to $S$ and $\mbf{y}_K=\mbf{y}$. The marginal model of $\mbf{y}_k$ is an ARMA model that depends on a parameter subset with a partition $\bsy\theta_k,\bsy\eta_k$ where 
    $$
    \bsy\theta_k=\mu_k\cup\sigma_k\cup\left(\bigcup_{i=1}^p\phi_{i,k}\right)\cup\left(\bigcup_{j=1}^q\pi_{j,k}\right)
    $$
    and $\bsy\eta_k=\varnothing$ for $k$ from 1 to $S$. The proof is provided in Appendix \ref{apdx:dVARMA}. Note that this model is MP as a result of the AR and MA parameter matrices being diagonal. The marginal (joint) model of $\mbf{y}_K$ is a diagonal VARMA model that depends on a parameter subset with a partition $\bsy\theta_K,\bsy\eta_K$ where $\bsy\theta_K=\bsy\nu$ and $\bsy\eta_K=\bsy\theta_1\cup\dots\cup\bsy\theta_S$. Clearly, this finite sequence of data subsets satisfies the conditions of Definition~\ref{def:MP}. 

\section{\label{sec:inference}Inference}
This section introduces the SMLE method for estimating the parameter set of a MP model in multiple steps, the SMLE consistency theorem and the details of SMLE for the diagonal VARMA model. 

\subsection{\label{ssec:SMLE}Stepwise Maximum Likelihood Estimation}
Assume that $\mcl{L}(\bsy\theta\mid\mbf{y})$ is a MP model likelihood function for $\mbf{y}$ and denote $(\mbf{y}_k)$ the corresponding finite sequence of data subsets that satisfies Definition~\ref{def:MP}. Let $\mcl{L}_k(\bsy\theta_k,\bsy\eta_k\mid\mbf{y}_k)$ denote the marginal likelihood function of $\mbf{y}_k$, that depends on a parameter subset with partition $\bsy\theta_k,\bsy\eta_k$ for $k=1,\dots,K$. Instead of estimating $\bsy\theta$ with the MP joint likelihood function $\mcl{L}(\bsy\theta\mid\mbf{y})$ in one step (MLE), the SMLE method estimates $\bsy\theta_1$ with the marginal likelihood function $\mcl{L}_1(\bsy\theta_1\mid\mbf{y})$ in step one (since $\bsy\eta_1=\varnothing$) and estimates $\bsy\theta_k$ with the marginal estimated likelihood function $\widehat{\mcl{L}}_k(\bsy\theta_k\mid\mbf{y})=\mcl{L}_k(\bsy\theta_k,\widehat{\bsy\eta}_k\mid\mbf{y})$ in step $k=2,\dots,K$. Here $\widehat{\bsy\eta}_k$ is obtained from primary parameter estimates obtained in previous steps. The SMLE is detailed in Algorithm 1.

\begin{algorithm}
\setstretch{1.4}
\caption{Sequential Maximum Likelihood Estimation}
\label{alg:SMLE}
\begin{algorithmic}[1]
\State $\widehat{\bsy\theta}_1\gets\text{arg~max}_{\boldsymbol\theta_1}~\mcl{L}_1(\boldsymbol\theta_1\mid\mbf{y})$
\For{$k\gets2~to~K$}
\label{opt} \State $\widehat{\bsy\theta}_k\gets\text{arg~max}_{\boldsymbol\theta_k}~\widehat{\mcl{L}}_k(\boldsymbol\theta_k\mid\mbf{y})$
\EndFor
\State \textbf{return} $\widehat{\bsy\theta}=\widehat{\bsy\theta}_1\cup\dots\cup\widehat{\bsy\theta}_K$
\end{algorithmic}
\end{algorithm}

In general this is a sequential algorithm; under certain conditions, however, sequences of steps can be parallelized and performed in one \textit{stage}. For example, the steps from $k+1$ to $k+n$ can be parallelized if $\widehat{\bsy\eta}_{k+j}$ for $j=1,\dots,n$ can be obtained after step $k$, since all the marginal estimated likelihood functions can be obtained after step $k$. Formally, the steps from $k+1$ to $k+n$ can be performed independently if $\bsy\eta_{k+j}\subseteq\bsy\theta_1\cup\dots\cup\bsy\theta_k$ for $j=1,\dots,n$. Note that the SMLE method can be performed in parallel when estimating the parameters of the ARMA models of the diagonal VARMA model, see Subsection~\ref{ssec:dARMAinf}. Hence, the parameters of the diagonal VARMA model can be obtained in two stages.

\subsection{\label{ssec:cons}Consistency}
We assume the data set $\mbf{y}$ is a realization from a MP model with true parameter set $\bsy\theta^*$ and denote $\mbf{Y}$ as the corresponding random vector. Define $\widehat{\bsy\theta}_1(\mbf{Y})$ as the estimator of $\bsy\theta_1$, define $\widehat{\bsy\theta}_k(\mbf{Y},\bsy{\eta}_k)$ as the estimator of $\bsy\theta_k$ given $\bsy{\eta}_k$, define $\widehat{\bsy\theta}_k{'}(\mbf{Y},\bsy{\eta}_k)$---if it exists---as its Jacobian matrix (derivative with respect to $\bsy{\eta}_k$) and define $\widehat{\bsy{\eta}}_k(\mbf{Y})$ as the estimator of $\bsy\eta_k$. Furthermore, let $n_k$ quantify the information contained in $\mbf{Y}$ relevant to the estimation of $\bsy{\theta}_k$. For the diagonal VARMA model $n_k=T$ for $k$ from $1$ to $S$ and $n_K=S$.  The SMLE consistency theorem provides the conditions under which $\widehat{\bsy{\theta}}_k(\mbf{Y},\widehat{\bsy{\eta}}_k(\mbf{Y}))$ is consistent for $k=2,\dots,K$.

\begin{theorem}[SMLE Consistency]
\label{thrm:cons}
Suppose that
\begin{eqnarray}
\label{ass_one}
\widehat{\bsy{\theta}}_1(\mbf{Y})\overset{P}{\longrightarrow}\bsy{\theta}_1^*&\text{as}&n_1\rightarrow\infty,\\
\label{ass_two}
\widehat{\bsy{\theta}}_k(\mbf{Y},\bsy{\eta}_k^*)\overset{P}{\longrightarrow}\bsy{\theta}_k^*&\text{as}&n_k\rightarrow\infty,
\end{eqnarray}
for $k=2,\dots,K$ where $\bsy{\eta}_k^*$ and $\bsy{\theta}_k^*$ are the true parameter sets for all $k$. Furthermore, assume that there exists a $n_{k0}<\infty$ such that for all $n_k>n_{k0}$, $\widehat{\bsy{\theta}}_k{'}(\mbf{Y},\bsy\eta_k)$ exists and is uniformly bounded in an open neighborhood of $\bsy{\eta}_k^*$ almost surely for $k=2,\dots,K$. Then
\begin{equation*}
\widehat{\bsy{\theta}}_k(\mbf{Y},\widehat{\bsy{\eta}}_k(\mbf{Y}))\overset{P}{\longrightarrow}\bsy{\theta}_k^*~~~\text{as}~~~n_1,\dots,n_k\rightarrow\infty,
\end{equation*}
for $k=2,\dots,K$.
\end{theorem}

The assumptions of the SMLE consistency theorem are required for the Spall consistency theorem \citep[Theorem~1]{spall1989effect} used in the inductive hypothesis of the proof provided in appendix \ref{apdx:cons}. Heuristically, for step $k=2,\dots,K$ the theorem states that if $\widehat{\bsy\eta}_k$ is a consistent estimator, $\widehat{\bsy\theta}_k(\mbf{Y},\bsy\eta_k^*)$ is a consistent estimator and $\widehat{\bsy{\theta}}_k{'}(\mbf{Y},\bsy\eta_k)$ exists and is well-behaved near $\bsy\eta_k^*$, then $\widehat{\bsy\theta}_k(\mbf{Y},\widehat{\bsy\eta}_k)$ is a consistent estimator.

\subsection{\label{ssec:dARMAinf}Diagonal VARMA Inference}
As $\bsy\eta_k=\varnothing$ for $k$ from $1$ to $S$, the parameters $\bsy\theta_k$ for $k$ from $1$ to $S$ can be estimated in parallel with ARMA likelihood functions in one stage. The temporal parameter estimates obtained in this stage are consistent in $T$ \citep[][Section~5.8]{hamilton1994time} and satisfy assumptions \eqref{ass_one} and \eqref{ass_two} of the SMLE consistency theorem \eqref{thrm:cons} where $n_k=T$ for $k$ from 1 to $S$. The step $K$ marginal (joint) likelihood function has an innovation form \citep{schweppe1965evaluation}
\begin{equation}
\label{eq:innform}
\mathcal{L}_{K}(\boldsymbol\theta_{K},\boldsymbol\eta_{K}\mid\mathbf{y})=\prod_{t=1}^{T}g(\boldsymbol\theta_{K}\mid\mathbf{u}_{t}(\boldsymbol\eta_{K}\mid\mathbf{y})),
\end{equation}
where $g(\cdot)$ is the innovation likelihood function and $\mathbf{u}_{t}(\boldsymbol\eta_{K}\mid\mathbf{y})$ are the $S$ residuals of the data set at time point $t$ that depend on $\boldsymbol\eta_{K}$. In stage two the spatial parameter estimates are obtained from \eqref{eq:innform} using the residuals obtained in stage one. Hence, this is a two-stage approach. This approach has some advantages. First, ARMA model selections can be performed in the first stage, i.e.~the AR and MA orders of the ARMA models can vary across space. Second, innovation model selection and specification can be performed in the second stage. Note that if the innovation model is a MP model and the parameter set can be estimated in two stages, then the parameter set of the diagonal VARMA model can be estimated in three stages, see the application in Section~\ref{sec:app} for an example.

\section{\label{sec:sim}Simulation Study}
	The first simulation study compares the small sample biases and standard errors of the maximum likelihood and stepwise maximum likelihood estimators for the diagonal VARMA model \eqref{eq:dVARMA} introduced in Subsection \ref{ssec:dVARMA} with isotropic innovations. The second simulation study is used to corroborate the SMLE consistency Theorem \ref{thrm:cons} introduced in Section \ref{ssec:cons} for the same model.

\subsection{\label{ssec:simmod}Simulation Model}
	The diagonal VARMA model \eqref{eq:dVARMA} is used with zero mean, AR order two and MA order zero, i.e.~centered diagonal VAR(2), with isotropic innovations. The number of parameters is restricted so that the SMLE method can be directly compared with the MLE method where the number of evaluations grows exponentially with the number of parameters. For all $T$ and $S$ considered in this section define the standard deviation and AR parameters as $\sigma_s=1.2$, $\phi_{1,s}=0.50$ and $\phi_{2,s}=0.25$ for all $s$. The isotropic innovations are modeled with the Mat\'{e}rn correlation function, which for distance $\mathbf{h}$ has the following form: $\frac{\pi^{1/2}}{2^{\kappa-1}\Gamma(\kappa+1/2)\alpha^{2\kappa}}\left(\alpha\lVert \mathbf{h}\lVert\right)^\kappa K_\kappa(\alpha\lVert \mathbf{h}\lVert)$,
where the inverse scale parameter $\alpha>0$ controls the range of correlation, the smoothness parameter $\kappa>0$ controls the mean-square differentiability of the process and $K_\kappa(\cdot)$ is a modified Bessel function \citep[p.~31]{stein2012interpolation}. In this section we define the inverse scale and smoothness parameters as $\alpha=0.3$ and $\kappa=1.5$ respectively. This model corresponds to $N = T\cdot S$ data points, where spatial locations are distributed regularly on a line, and $P=3S+2$ parameters, i.e.~three temporal parameters for each time series and two spatial parameters.

\subsection{Set-up}
\label{setup}
	Since, for MLE, the number of evaluations grows exponentially with the number of parameters and $P=3S+2$, the number of spatial locations $S$ is restricted so that the SMLE method can be directly compared to the MLE method. Therefore, for the purpose of comparison, we let $T=50$ and $S=20$ (so that $N=50\cdot 20=1,000$ and $P=3\cdot20+2=62$). For both methods the algorithm is initialized at the true parameter values to eliminate the effects of initial value selection and standard errors are obtained by performing both the MLE and SMLE methods for 30 independent simulations. All numerical optimizations (for MLE and SMLE) are performed in \texttt{R} with the Nelder-Mead simplex algorithm \citep{nelder1965simplex}. For SMLE the first $S$ steps are performed in parallel via hyper-threading with a 2.6 GHz Intel Core i7 (8 virtual cores) CPU and 8 GB of RAM.

\subsection{Results}
	Table~\ref{tab:ests} displays the mean estimates (averaged over $s$) and standard errors, in parenthesis, of $\sigma_s$, $\phi_{1,s}$, $\phi_{2,s}$, $\alpha$ and $\kappa$ obtained from SMLE, Full MLE and Fixed MLE ($\sigma_s$ fixed for all $s$) using 30 independent simulations. The estimates from Full MLE suggest that $\sigma_s$ for all $s$ are not identifiable in the diagonal VARMA model but are identifiable in the ARMA models. Consequently, Fixed MLE was included to provide a comparison with SMLE. The mean estimates and standard errors obtained from Fixed MLE aim to approximate those from Full MLE. The approximated estimates are expected to be more statistically efficient as fixing the standard deviation parameters has increased the data to parameter ratio. There is a relatively small difference in the sample biases of $\widehat{\phi}_{1,s}$ and $\widehat{\phi}_{2,s}$ between SMLE and Fixed MLE, however, the 15\% and 14\% relative efficiencies (ratio of standard errors), respectively, demonstrate that the estimates from SMLE are less efficient, as expected. The sample biases of $\widehat{\alpha}$ and $\widehat{\kappa}$ are relatively small with relative efficiencies 50\% and 57\% respectively. Considering that these are conservative estimates and the number of data points is small ($N=1,000$) there appear to be relatively small sample biases. The fact that SMLE was capable of estimating $\sigma_s$ for all $s$ demonstrates one advantage of estimating with marginal likelihood functions.

\begin{table}[ht]
\caption{Mean estimates (averaged over $s$) and standard errors, in parenthesis, of $\sigma_s$, $\phi_{1,s}$, $\phi_{2,s}$, $\alpha$ and $\kappa$ obtained from SMLE, Full MLE and Fixed MLE using 30 independent simulations.}
\begin{center}
	\begin{spacing}{1.5}
		\begin{tabular}{r c c c c c}
			\hline
			& $\widehat{\sigma}$ & $\widehat{\phi}_{1}$ & $\widehat{\phi}_{2}$ & $\widehat{\alpha}$ & $\widehat{\kappa}$ \\
			\hline\hline
			\textbf{SMLE} & 1.16 (0.12) & 0.51 (0.13) & 0.21 (0.14) & 0.33 (0.04) & 1.54 (0.07) \\
			\textbf{Full MLE} & 95.13 (1746.27) & 0.30 (0.59) & -0.03 (0.52) & 0.15 (0.11) & 1.27 (0.40) \\
			\textbf{Fixed MLE} & NA & 0.50 (0.02) & 0.24 (0.02) & 0.30 (0.02) & 1.51 (0.04) \\
			\hline
			True Values & 1.20 & 0.50 & 0.25 & 0.30 & 1.50 \\
			\hline
		\end{tabular}
		\label{tab:ests}
	\end{spacing}
\end{center}
\end{table}

	Table~\ref{timeIter} displays the mean time (seconds) and mean number of iterations required by both estimation methods and the two steps of the SMLE method. On average MLE took approximately 4,800 times longer (24 minutes 4 seconds) and required approximately 700 times more iterations than SMLE. The relative errors of the times and iterations are $6\%$ and $10\%$, respectively, for SMLE and $34\%$ and $24\%$, respectively, for MLE. \bigskip
\begin{table}[ht]
\caption{Mean time (seconds) and mean number of iterations required by both estimating methods and the two steps of the SMLE method. Means and standard deviations, in parenthesis, are calculated from the 30 simulations.}
\begin{center}
	\begin{spacing}{1.5}
		\begin{tabular}{c c c c | c}
			\hline
		    & & \textbf{SMLE} & & \textbf{MLE}\\
			& Step 1 & Step 2 & Total & Total \\
			\hline\hline
			Time (sec) & 0.09 (0.00) & 0.20 (0.02) &  0.30 (0.02) & 1444.17 (486.67)\\
			\# Iterations & 125.73 (18.12) & 50.60 (5.72) & 176.33 (18.27) & 126178.30 (30820.19)\\
			\hline
		\end{tabular}
		\label{timeIter}
	\end{spacing}
\end{center}
\end{table}

\begin{figure}[H]
	\begin{center}
		\includegraphics[width=15cm]{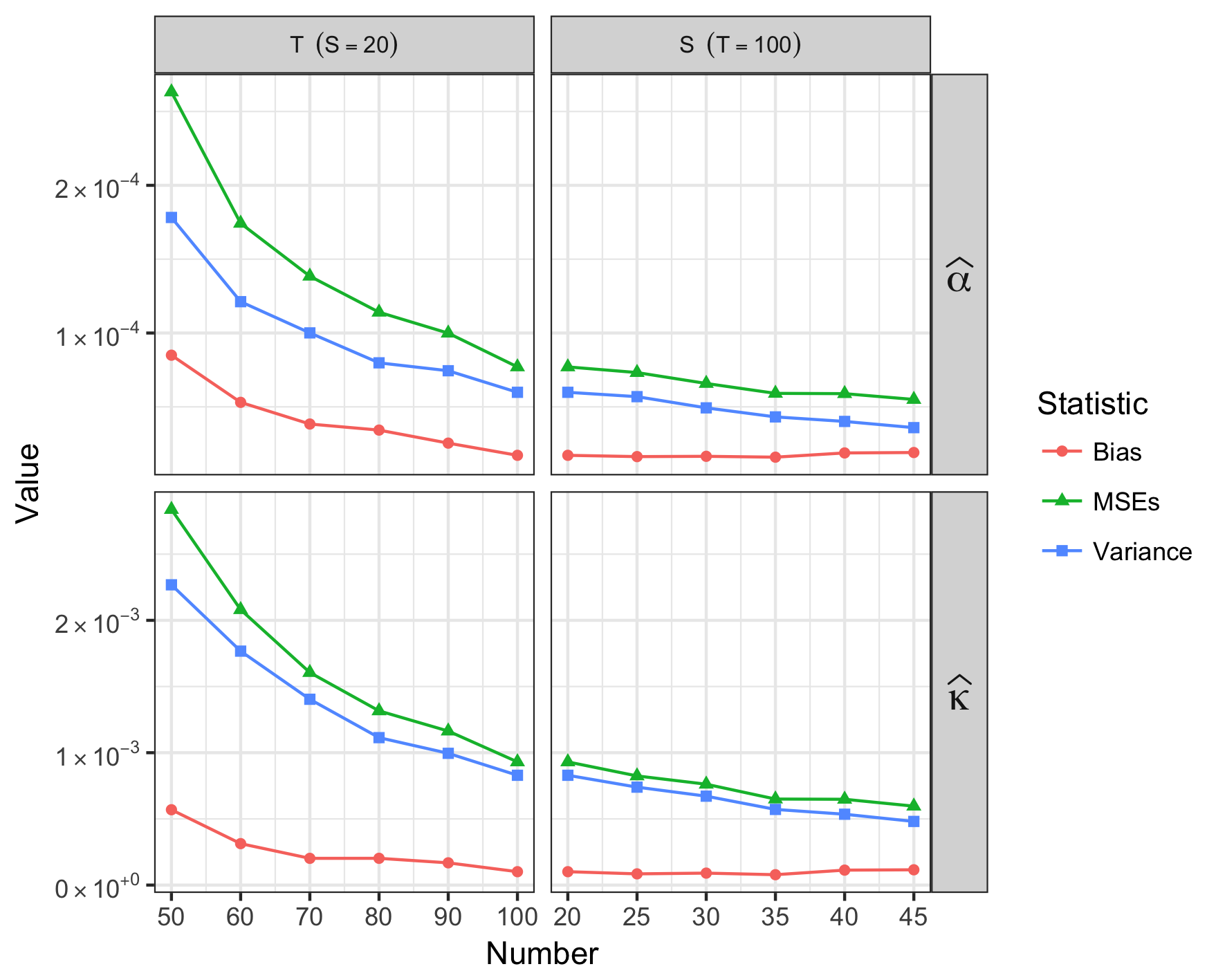}
  		\caption{Bias (squared), variance and mean square error (MSE) of $\widehat{\alpha}$ and $\widehat{\kappa}$ from SMLE as the number of time points $T$ increases to 100 with $S=20$ fixed and then as the number of spatial locations $S$ increases to 45 with $T=100$ fixed.}
  		\label{MSE}
	\end{center}
\end{figure}

	Estimators are expected to have zero bias and variance as $T$ and $S$ increase to infinity according to the consistency theorem. However, for clarity we provide plots of the bias, variance and MSE of estimators as $T$ increases with $S$ fixed then as $S$ increases with $T$ fixed. Figure \ref{MSE} displays the bias (squared), variance and MSE profile plots of $\widehat{\alpha}$ and $\widehat{\kappa}$ from SMLE as the number of time points $T$ increases to 100 with $S=20$ fixed and then as the number of spatial locations $S$ increases to 45 with $T=100$ fixed. These plots provide numerical evidence of the SMLE consistency theorem in Subsection \ref{ssec:cons} as they demonstrate how the MSE decreases for $\widehat{\alpha}$ and $\widehat{\kappa}$ conditional on the temporal parameter estimates as $T$ increases for a fixed number of spatial locations ($S=20$) and as $S$ increases for a fixed number of time points ($T=100$). For both $\widehat{\alpha}$ and $\widehat{\kappa}$ the variance dominates the bias for all values of $T$. Furthermore, the plots demonstrate that the reductions in MSE are mostly attributable to reductions in variance with small bias even for a few time points and spatial locations.

\section{\label{sec:app}Application}

\subsection{Data}
\label{ssec:data}
To illustrate the flexibility of a MP model and the efficiency of SMLE for large data we consider the large ensemble community project data set \citep{kay2015community}. This data set is an ensemble of Earth System Model simulations, consisting of hundreds of spatio-temporal variables. From this data set a single member of the near surface (2 m above surface) temperature variable is selected. This variable was simulated from 2006 to 2100 according to the Representative Concentration Pathways 8.5 scenario \citep{van2011representative}, $T=95$ years, over a discrete global grid consisting of $N=288$ longitudes and $M=192$ latitudes. The temperature variable was annually averaged to approximate, through the central limit theorem, the Gaussian assumption of the diagonal VARMA model (diagnostics of the other assumption is provided in \cite{cas18}), for a total of $T\times N\times M=5,253,120$ data points.

\subsection{\label{ssec:app_mod}A Model for Global Data}
A diagonal VAR(2) model \eqref{eq:dVARMA} is applied, and innovations $\mathbf{U}_t$ are assumed stationary in longitude but nonstationary in latitude \cite[axial symmetry,][]{jones1963stochastic}. The vector of mean parameters as in \eqref{eq:dVARMA} is $\bsy\mu=\bsy\beta_0+t\bsy\beta_1$ where $\bsy\beta_j=(\beta_{1,j},\dots,\beta_{S,j})$ for $j=1, 2$. 
Given the discrete geometry of the data and the axial symmetry assumption, the correlation matrix $\text{R}(\bsy\nu)$ of $\mathbf{U}_t$ in \eqref{eq:dVARMA} is block-circulant \citep{davis79circulant}, so the $M\times M$ blocks are circulant with the form
\begin{equation*}
\text{R}_{m_1,m_2}=\text{C}\text{F}_{m_1,m_2}\text{C}\hrmt, \quad m_1, m_2= 1, \ldots, M,
\end{equation*}
where $\text{C}=\mcl{C}/\sqrt{N}$ is a unitary matrix, $\mcl{C}$ is the unnormalized $N\times N$ discrete Fourier transform (DFT) matrix and $\text{F}_{m_1,m_2}=\text{diag}(\mbf{f}_{m_1,m_2})$ is a diagonal matrix \citep[][Section~3.2]{davis79circulant}. The elements of $\mbf{f}_{m_1,m_2}$ are the values of the cross-spectral mass function, denoted $f_{m_1,m_2}(\cdot)$, between latitudes $L_{m_1}$ and $L_{m_2}$. Hence, block-circulant matrices are modeled in the spectral domain. The cross-spectral mass functions have the form
\begin{equation}
\label{csmf}
f_{m_1,m_2}(c)=f_{m_1}^{1/2}(c)\cdot f_{m_2}^{1/2}(c)\cdot\rho_{m_1,m_2}(c),
\end{equation}
where $f_{m}(\cdot)$ is the spectral mass function at latitude $L_m$ and $\rho_{m_1,m_2}(\cdot)$ is the coherence function between latitudes $L_{m_1}$ and $L_{m_2}$.

	Following \cite{castruccio2013global, poppick2014using} the spectral mass function for the innovations at latitude $L_m$ is the modified Mat\'{e}rn defined as
\begin{equation*}
f_m(c)\propto\frac{1}{\left(\alpha_m^2+4\sin^2\frac{\pi c}{N}\right)^{\kappa_m+1/2}}\quad\text{s.t.}\quad\frac{1}{N}\sum_{c=0}^{N-1}f_m(c)=1,
\end{equation*}
where $\alpha_m$ is the inverse range parameter and $\kappa_m$  controls the increased decay rate in spectral mass for larger wavenumbers. The normalization condition is required as the innovations are unscaled. The coherence function is defined as 
\begin{equation*}
\rho_{m_1,m_2}(c)\coloneqq\left[\frac{\xi}{(1+4\sin^2\frac{\pi c}{N})^\tau}\right]^{|L_{m_1}-L_{m_2}|},
\end{equation*}
where $\xi$ controls the rate of decay in coherence, over all wavenumbers, as the distance between latitudes increases and $\tau$ controls the increased decay rate in coherence for larger wavenumbers. Therefore, the correlation matrix $\text{R}(\bsy\nu)$ has the parameter set
\begin{equation*}
\bsy\nu=\left(\bigcup_{m=1}^M\alpha_m\right)\cup\left(\bigcup_{m=1}^M\kappa_m\right)\cup \xi \cup \tau.
\end{equation*}

\subsection{Estimation}
\label{ssec:est}

The model presented for the innovations $\mathbf{U}_t$ in the previous section is a MP model, and $\bsy\nu$ can be estimated in two stages. Consider the finite sequence of $K=M+1$ data subsets $(\mathbf{u}_m)$ where $\mathbf{u}_m$ are the innovations at a fixed latitude $L_m$ for $m=1,\dots,M$ across longitudes (a circle), and $\mbf{u}_K=\mathbf{u}$. The marginal model of $\mbf{u}_m$ is a Whittle model \citep{whittle1954stationary} that depends on a parameter subset with a partition $\bsy\theta_m,\bsy\eta_m$ where $\bsy\theta_m=\{\alpha_m,\kappa_m\}$ and $\bsy\eta_m=\varnothing$, the proof is provided in Appendix~\ref{apdx:axial}. The marginal (joint) model for $\mbf{u}_K$ depends on a parameter subset with a partition $\bsy\theta_K,\bsy\eta_K$ where $\bsy\theta_K=\{\zeta,\tau\}$ and $\bsy\eta_K=\bsy\theta_1\cup\dots\cup\bsy\theta_M$. Clearly, this finite sequence of data subsets satisfies the conditions of Definition~\ref{def:MP}. Consequently, the parameter set of the diagonal VARMA model can be estimated in three stages rather than two in the case of an unstructured correlation matrix $R(\bsy\nu)$. In the first stage the temporal diagonal VAR parameters are estimated, in the second stage the longitudinal spectral mass function parameters are estimated and in the third stage the latitudinal coherence function parameters are estimated.

\begin{figure}[H]
	\begin{center}
		\includegraphics[width=8cm]{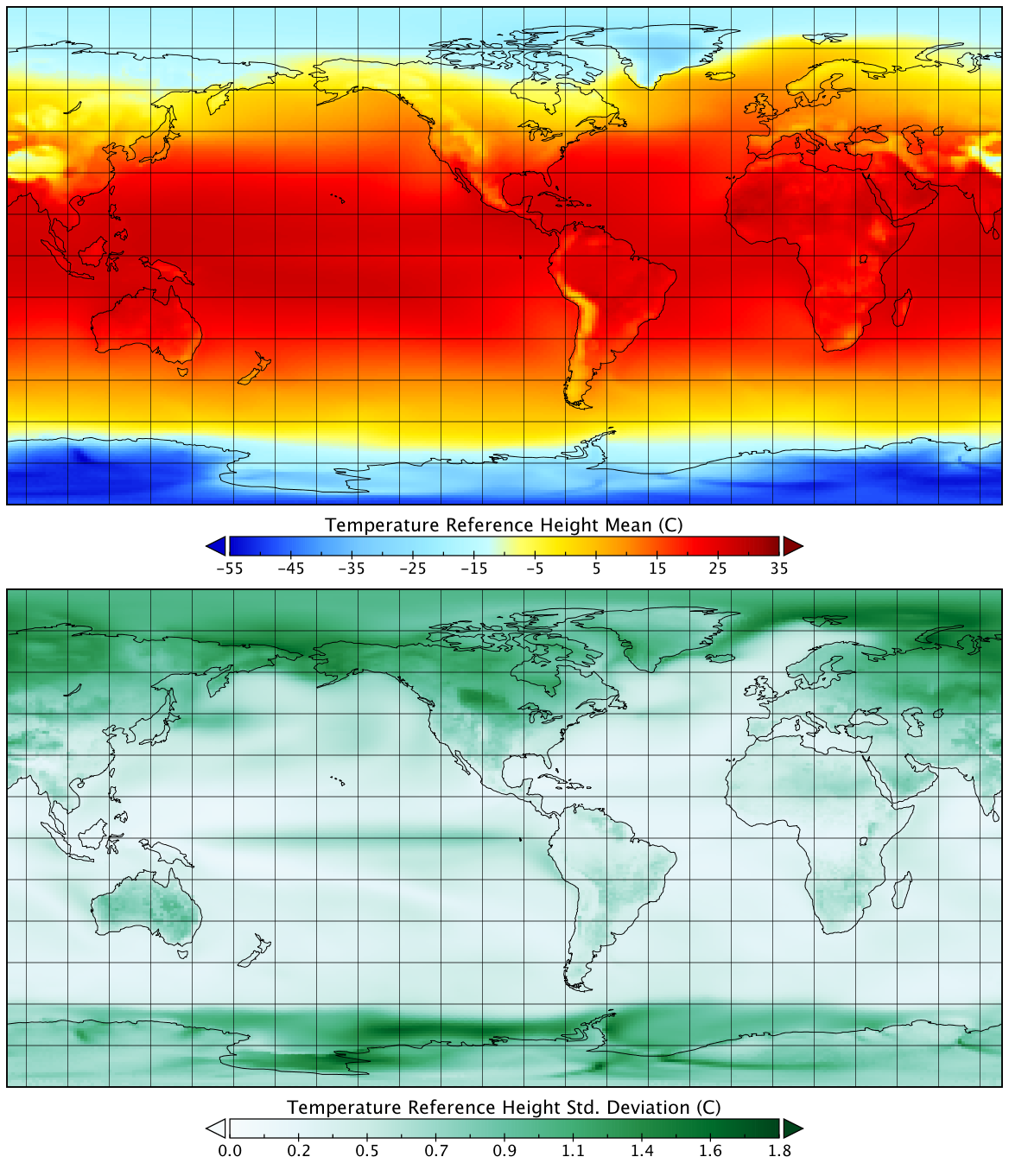}\hfill 
        \includegraphics[width=8cm]{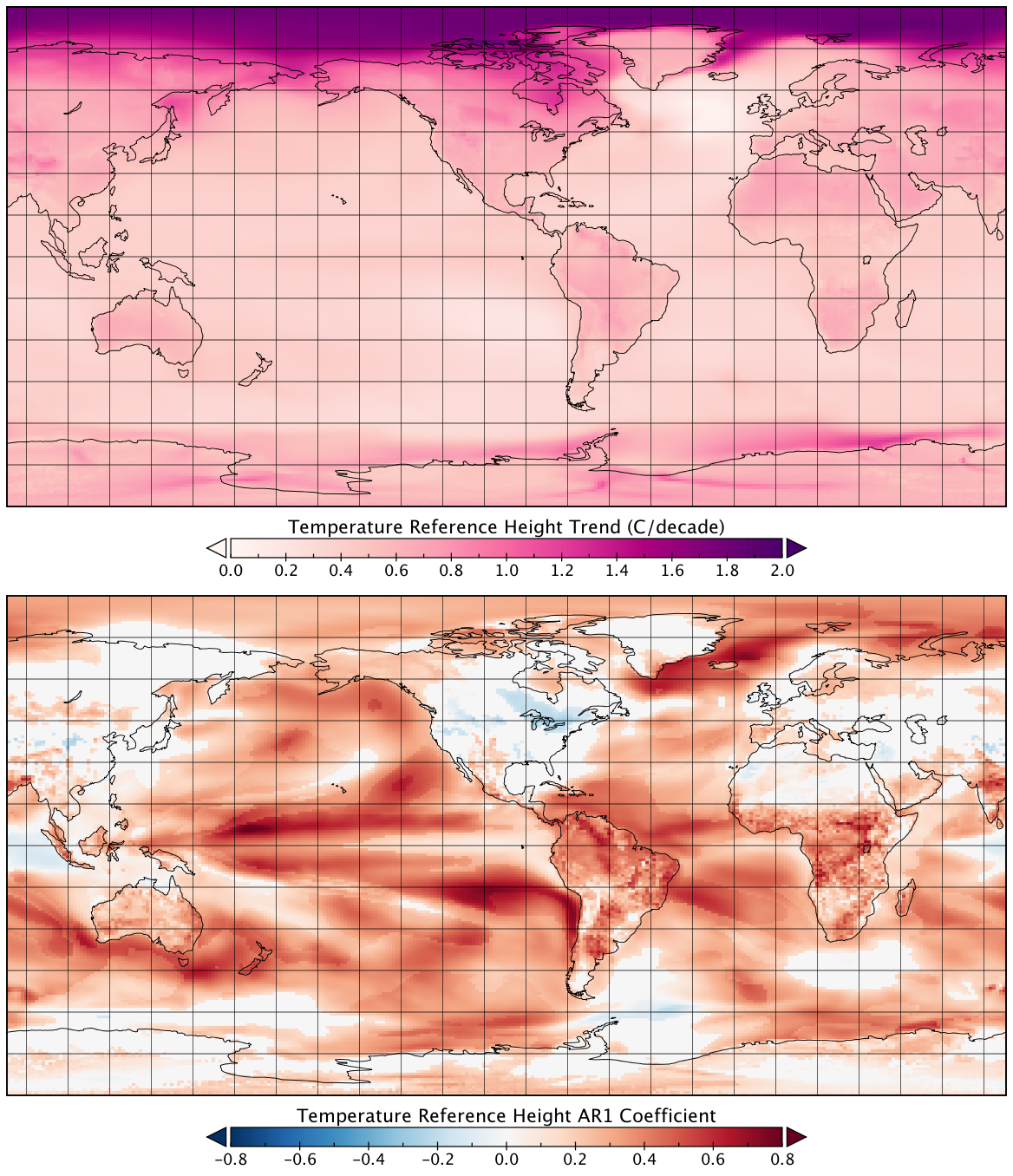}
  		\caption{The mean (upper left), standard deviation (lower left), trend (upper right) and order one AR coefficient (lower right) parameters of the marginal AR models over the discrete global grid.}
  		\label{fig:esm}
	\end{center}
\end{figure}

	Figures~\ref{fig:esm} displays the mean, standard deviation, trend, and order one AR coefficient parameters of the marginal AR models over the discrete global grid. The mean parameters capture the variation in temperature that results from latitude and altitude (e.g.~Himalayas, Andes). The standard deviation parameters capture the difference in temperature variability over land surface and sea. The trend parameters capture the lower rates of increase of temperature over the sea. The order one autoregressive parameters capture the long range dependence over the equatorial Pacific ocean due to irregular periodic variation of the El Ni\~no Southern Oscillation.

\begin{figure}[H]
	\begin{center}
		\includegraphics[width=16cm]{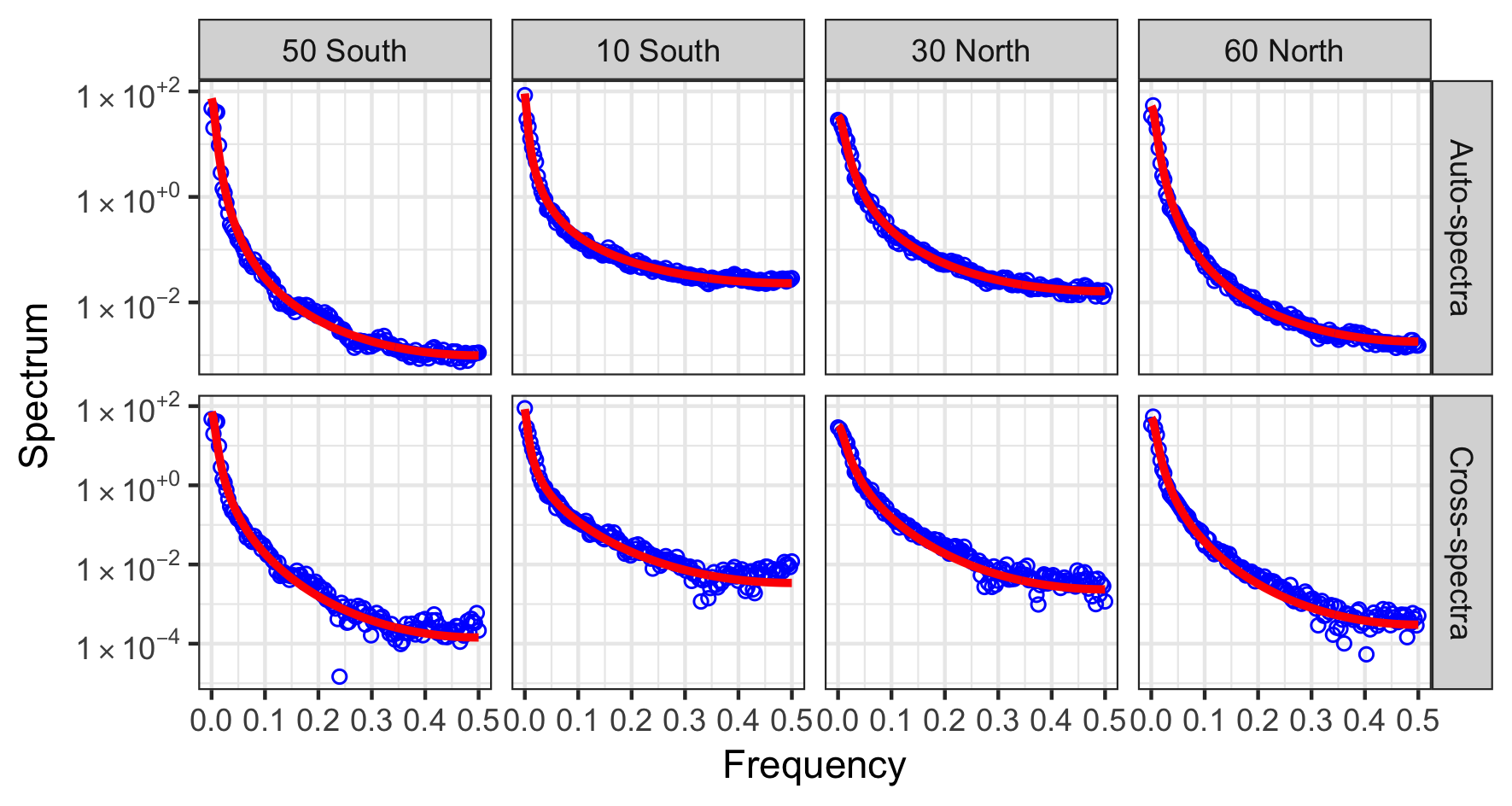}
  		\caption{(top row) The periodogram (blue) and spectral mass functions (red) and (bottom row) the cross-periodogram (blue) and cross-spectral mass function (red) at four latitudes.}
  		\label{fig:long_pars}
	\end{center}
\end{figure}

    Figure~\ref{fig:long_pars} (top row) displays periodograms (blue) and fitted spectral mass functions (red) at four selected latitudes. Figure~\ref{fig:long_pars} (bottom row) displays cross-periodograms (blue) and fitted cross-spectral mass functions (red) at the corresponding latitudes. These figures demonstrate that the spectral mass functions and the coherence function have provided a reasonable model for the correlations within and between, respectively, the data at each latitude given the axially symmetric assumptions.

	Estimation was performed on the laptop computer described in Subsection~\ref{setup}. Table~\ref{tab:times} displays the time required by the SMLE algorithm to estimate the temporal, longitudinal and latitudinal parameters. This table also includes the number of parameters and data points corresponding to each marginal model for the three stages. In total, this model has been fit to over five million data points in forty minutes. This model can be fit to larger data sets if a cluster of computers is available. However, the size and complexity of this data set demonstrates that fitting a diagonal VARMA model to a large complex spatio-temporal data set is possible on a laptop computer.
\begin{table}
\begin{center}
    \begin{spacing}{1.5}
		\begin{tabular}{c c c c c}
			\hline
			Stage & Time & \# Parameters / Marginal Model & \# Data points / Marginal Model \\
			\hline\hline
			Temporal & 9 mins & 4 & 95 \\
            Longitudinal & 7 secs & 2 & 27,360 \\
            Latitudinal & 31 mins & 3 & 55,253,120 \\
			\hline
		\end{tabular}
        \caption{The time required by the SMLE algorithm to estimate the temporal, longitudinal and latitudinal parameters, and the number of parameters and data points corresponding to each submodel for the three stages.}
    	\label{tab:times}
	\end{spacing}
    
\end{center}
\end{table}

\section{\label{sec:con}Conclusion and Future Work}

This work lays the foundation of a new approach for modeling large and complex spatio-temporal data, by predicating a partition of the parameter space in order to achieve inference in a multi-step fashion, the key principle being that the dependence of some data subsets can be described exclusively by some parameters in the model.    

MP models are very broad and flexible, and can be efficiently applied to large complex data sets via SMLE. The SMLE method can be parallelized, and we provide conditions under which the estimators are consistent. We demonstrate that a diagonal VARMA is MP, and that the parameters of this model can be estimated at least three orders of magnitude faster using SMLE compared to MLE, with very little loss in statistical efficiency. A MP model was applied to a large complex global climate data set consisting of over five million data points with a laptop computer in forty minutes. The model demonstrated the capacity to capture important complex features such as spatially varying means and nonstationary global dependence.

	Proposing a MP model for a particular application requires the existence of a sequence of data subsets that satisfies the definition, and while in our application this was suggested by the geometry of the problem, an automatic approach comprising of a clustering method to identify this sequence would be desirable. It would also be desirable to develop a selection of MP models, for different applications, that satisfies the assumptions of the SMLE consistency theorem.
    
    The parameter set of a MP model can be estimated efficiently via SMLE. This is a frequentist method that depends on the assumption that parameter subsets can be separated from the joint model with marginal models and estimated with marginal likelihood functions. Since Bayesian methods have been applied to composite marginal likelihood functions \citep{pauli2011bayesian} there could be the opportunity to embed MP models in tools of modern Bayesian inference. 
    
    The first simulation study suggested that the standard deviation parameters are not identifiable in the diagonal VARMA model but are identifiable in the ARMA models. In general, it is important to know if the identifiability of a set of parameters can change with respect to a marginal model. If the identifiability of a set of parameters can change with respect to a marginal model, it is still of interest to understand if this is the case for the diagonal VARMA model or whether these parameters are just very difficult to estimate.
    
\bibliographystyle{agsm}
\bibliography{library}

\begin{appendices}
\section{Marginal Diagonal VARMA Model}
\label{apdx:dVARMA}
\begin{proof}
If the diagonal VARMA model \eqref{eq:dVARMA} is stable then it can be represented as
$\mathbf{Y}_t=\boldsymbol{\mu}+\sum_{i=0}^\infty\Psi_i\Sigma\mathbf{U}_{t-i}$ \citep[p.~421]{lutkepohl2005new} where $\Psi_i$ is a linear combination of the diagonal matrices of AR and MA parameters. Consequently, $\Psi_i=\text{diag}(\psi_{i,1},\dots,\psi_{i,S})$ are the diagonal matrices of infinite order MA parameters. The mean vector and lag-$h$ autocovariance matrix function of this model are $\boldsymbol\mu$ and 
\begin{equation}
\Gamma(h):=\sum_{i=0}^\infty\Psi_{i+h}\Sigma\text{R}(\bsy\nu)\Sigma\Psi_{i}
\end{equation}
respectively. Therefore, the mean and lag-$h$ autocovariance function of the marginal model of $\mbf{y}_k$ is $\mu_k$ and $\gamma_k(h):=\sigma_k^2\sum_{i=0}^\infty\psi_{i+h,k}\psi_{i,k}$. The marginal model of $\mbf{y}_k$ is therefore an ARMA model that depends on a parameter subset with a partition $\bsy\theta_k,\bsy\eta_k$ where 
	$$
    \bsy\theta_k=\mu_k\cup\sigma_k\cup\left(\bigcup_{i=1}^\infty\psi_{i,k}\right)=\mu_k\cup\sigma_k\cup\left(\bigcup_{i=1}^p\phi_{i,k}\right)\cup\left(\bigcup_{j=1}^q\pi_{j,k}\right)
    $$
and $\bsy\eta_k=\varnothing$ for $k=1,\dots,S$.
\end{proof}

\section{SMLE Consistency}
\label{apdx:cons}
\begin{proof}
Base case $(k=2)$: Since $\widehat{\boldsymbol\eta}_2(\mathbf{Y})\subseteq\widehat{\boldsymbol{\theta}}_1(\mathbf{Y})$ by definition, $\widehat{\boldsymbol\eta}_2(\mathbf{Y})$ is consistent by assumption and $\widehat{\boldsymbol{\theta}}_2(\mathbf{Y},\widehat{\boldsymbol\eta}_2(\mathbf{Y}))$ is consistent by the Spall consistency theorem \citep{spall1989effect}. Inductive hypothesis $(k>2)$: Suppose the theorem holds for $k=2,\dots,n<K$. Since $\widehat{\boldsymbol{\eta}}_{n+1}(\mathbf{Y})\subseteq\widehat{\boldsymbol{\theta}}_1(\mathbf{Y})\cup\dots\cup\widehat{\boldsymbol{\theta}}_n(\mathbf{Y},\widehat{\boldsymbol\eta}_n(\mathbf{Y}))$ by definition, $\widehat{\boldsymbol{\eta}}_{n+1}(\mathbf{Y})$ is consistent by the inductive hypothesis and $\widehat{\boldsymbol{\theta}}_{n+1}(\mathbf{Y},\widehat{\boldsymbol\eta}_{n+1}(\mathbf{Y}))$ is consistent by the Spall consistency theorem \citep{spall1989effect}.
\end{proof}

\section{Marginal Axially Symmetric Model}
\label{apdx:axial}
\begin{proof}
The block of the block-circulant correlation matrix that corresponds to $\mbf{u}_k$ is the circulant matrix $\text{R}_{k,k}$. Since $\rho_{k,k}(c)=1$, the model has the following log probability density function
\begin{eqnarray*}
&&-\frac{1}{2}\ln\vert2\pi\text{R}_{k,k}\vert-\frac{1}{2}\mbf{u}_k\hrmt\text{R}_{k,k}^{-1}\mbf{u}_k\\
&=&-\frac{1}{2}\ln\vert2\pi\text{U}\text{F}_{k,k}\text{U}\hrmt\vert-\frac{1}{2}(\mcl{F}\mbf{u}_k)\hrmt\mbf{F}_{k,k}^{-1}\mcl{F}\mbf{u}_k\\
&=&-\frac{1}{2}\ln\vert2\pi\text{F}_{k,k}\vert-\frac{1}{2}\widetilde{\mbf{u}}_k\text{F}_{k,k}^{-1}\widetilde{\mbf{u}}_k\\
&=&-\frac{N}{2}\ln(2\pi)-\frac{1}{2}\sum_{c=0}^{N-1}\left(\ln f_{k}(c)-\frac{\vert\widetilde{\text{u}}_{k,c}\vert^2}{f_{k}(c)}\right)
\end{eqnarray*}
where $\widetilde{\text{u}}_{k,c}$ is an element of $\widetilde{\mbf{u}}_k$. The marginal model of $\mbf{u}_k$ is therefore a Whittle model \citep{whittle1954stationary} that depends on a parameter subset with a partition $\bsy\theta_k,\bsy\eta_k$ where $\bsy\theta_k=\{\alpha_k,\kappa_k\}$ and $\bsy\eta_k=\varnothing$ for $k=1,\dots,M$.
\end{proof}
\end{appendices}

\end{document}